\documentstyle[12pt]{article}
\newcommand{\doublespace}{
   \renewcommand{\baselinestretch}{1.5}
   \large\normalsize}

\def \Z{\Bbb Z}
\def \C{\Bbb C}

\def \Q{\Bbb Q}

\def \wt{{\rm wt}}

\def \End{{\rm End}}

\def \Ind {{\rm Ind}}
\def \Irr {{\rm Irr}}
\def \Aut{{\rm Aut}}
\def \Hom{{\rm Hom}}

\def \ann{{\rm Ann}}
\def \<{\langle}
\def \>{\rangle}

\def \a{\alpha }

\def \l{\lambda }

\def \o{\omega }
\def \c{\chi}

\def \cg{\chi_g}
\def \ag{\alpha_g}
\def \ah{\alpha_h}
\def \ph{\psi_h}
\def \nor{\vartriangleleft}
\input amssym.def
\doublespace
\input amssym
\begin{document}
\newtheorem{th1}{Theorem}
\newtheorem{thm}{Theorem}[section]
\newtheorem{prop}[thm]{Proposition}
\newtheorem{coro}[thm]{Corollary}
\newtheorem{lem}[thm]{Lemma}
\newtheorem{rem}[thm]{Remark}
\newtheorem{de}[thm]{Definition}
\newtheorem{hy}[thm]{Hypothesis}
\newcommand{\qed}{\mbox{ $\square$}}\newcommand{\pf}{\noindent {\em Proof: \,}}

\begin{center}
{\Large {\bf On the operator content of nilpotent orbifold models}} \\
\vspace{0.5cm}
Chongying Dong\footnote{Supported by NSF grant
DMS-9303374 and a research grant from the Committee on Research, UC Santa
Cruz.} \ \ \  and \ \ \ Geoffrey Mason\footnote{Supported by NSF grant
DMS-9122030 and a research grant from the Committee on Research, UC Santa
Cruz.}
\\
Department of Mathematics, University of
California, Santa Cruz, CA 95064
\end{center}

\hspace{1.5 cm}

\section{Introduction}
The present work is in essence a continuation of our paper [DM2]. To describe
our results we assume that the reader is familiar with the theory of vertex
operator algebras (VOA) and their representations (see for example [B], [FLM]
and [FHL]).

Suppose that $V$ is a {\em holomorphic} VOA and $G$ is a finite (and
faithful) group of automorphisms of $V.$ It is then a general
conjecture that the fixed vertex operator subalgebra $V^G$ of
$G$-invariants is {\em rational}, that is, $V^G$ has a finite number
of simple modules and every module for $V^G$ is completely
reducible. In fact, following the work of Dijkgraaf-Witten [DW] and
Dijkgraaf-Pasquier-Roche [DPR], one can formulate a precise conjecture
concerning the category Mod-$V^G$ of $V^G$-modules. Essentially
it says that Mod-$V^G$ is equivalent to the category of
$D^c(G)$-modules, where $D^c(G)$ is the so-called quantum double of
$G$ [Dr] modified by a certain 3-cocycle $c\in H^3(G,S^1).$ This
cocycle itself arises from a quasi-coassociative tensor product on
Mod-$D^c(G)$ which is expected to reflect the algebraic properties of
an appropriate notion of tensor product on the category Mod-$V^G.$

One of the goals of the present paper is to prove a variation
on this theme for a broad class of finite groups $G,$ not necessarily
abelian, under a suitable hypothesis concerning the so-called
{\em twisted sectors} for $V.$ Let us explain these results in more detail.

For each $g\in G$ we have the notion of a $g$-twisted sector, or
$g$-{\em twisted} $V$-{\em module}. It is an important conjecture, invariably
assumed in the physics literature, that there is a {\em unique} simple
$g$-twisted $V$-module (assuming that $V$ is holomorphic). This is known
in certain cases (eg. [D2] and [DM1]), but remains open in general. Let us
assume that it holds for now, and set
\begin{equation}\label{1.1}
V^*=\oplus_{g\in G}M(g)
\end{equation}
where $M(g)$ is the postulated unique simple $g$-twisted $V$-module.
In particular $M(1)=V$ where $1$ is the identity of $G.$ Our work is concerned
with an analysis of the sequence
\begin{equation}\label{1.2}
V^*\supseteqq V\supseteqq V^G.
 \end{equation}

It is straightforward to see that $V^*$ is a $V^G$-module, so we may
define the category ${\cal V}^*(G)$ to be the $V^G$-module category whose
objects are $V^G$-submodules of direct sums of copies of
$V^*,$ and whose morphisms are $V^G$-homomorphisms of
$V^G$-modules. As a special case of our results we have
\begin{th1}\label{t1} Let $V$ be a simple VOA and $G$ a faithful,
finite nilpotent group of
automorphisms of $V.$ Assume that for every $g\in G,$ there is
a unique simple $g$-twisted $V$-module. Then there is an equivalence
of categories
\begin{equation}\label{1.3}
\phi: {\cal V}^*(G)\to {\rm Mod-}D_{\a}(G).
\end{equation}
\end{th1}

We hasten to explain the notation. Following [DM2] and [M], if $C_G(g)=\{h\in
G|gh=hg\}$ is the {\em centralizer} of $g$ in $G$ then there is a {\em
projective representation} of $C_G(g)$ on $M(g)$ for each $g\in G.$
This data may be described by a certain $\a\in H^3(\Z G),$ the group
of Hochschild 3-cocycles on the integral group ring $\Z G.$ We can {\em
twist} the quantum double $D(G)$ by $\a$ to obtain another semi-simple
algebra $D_{\a}(G),$ and Theorem \ref{t1} identifies a certain category of
$V^G$-modules with the category Mod-$D_{\a}(G)$ of $D_{\a}(G)$-modules.
It remains to show that ${\cal V}^*(G)$ is the complete category of
$V^G$-modules, and that $D_{\a}(G)$ is the same as $D^c(G).$

Note that the origins of $D_{\a}(G)$ are quite different from those of the
algebra $D^c(G)$ mentioned above. But in addition to the fact that tensor
products of twisted modules are not presently understood, the algebra
$D_{\a}(G)$ is a very natural object to study in the present context. For we
will show that, quite generally, the space $V^*$ is naturally a module
over $D_{\a}(G),$ and that also $D_{\a}(G)$ commutes with action of the
vertex operators $Y(v,z)$ for $v\in V^G.$ Thus one may loosely say that
$V^*$ is a $D_{\a}(G)\otimes V^G$-module. On the other hand, we treat
$D_{\a}(G)$ solely as an associative algebra $-$ any quasi-quantum group
structure that is available is not relevant to the present considerations.

The precise nature of the equivalence $\phi$ of Theorem \ref{t1} is
perhaps as important as its existence. In case $G$ is nilpotent, we
will prove that $V^*$ decomposes into a direct sum
\begin{equation}\label{1.4}
V^*=\oplus_{\chi}M_{\chi}\otimes V_{\chi}
\end{equation}
where in (\ref{1.4}), $\chi$ ranges over the simple characters of
$D_{\a}(G),$ $M_{\chi}$ is a module over $D_{\a}(G)$ which affords $\chi,$ and
$V_{\chi}$ is a certain simple $V^G$-module. Then the map $\phi$ is (\ref{1.3})
is just the extension of a {\em bijection}
\begin{equation}\label{1.5}
\phi: M_{\chi}\to V_{\chi}
\end{equation}
{}from simple $D_{\a}(G)$-modules to simple $V^G$-modules which are contained
in $V^*.$

{}From this one can see that $D_{\a}(G)$ is precisely the {\em grade-preserving
commuting algebra} of $V^G$ on $V^*,$ that is $D_{\a}(G)$ is the
algebra of operators on $V^*$ which commute with the VOA $V^G$ and
which preserve the conformal grading on $V^*.$ Thus $D_{\a}(G)$ and
$V^G$ behave in many ways like a pair of mutually commuting algebras,
or a dual pair in representation theory. From this point
of view, the decomposition (\ref{1.4}) and
bijective correspondence (\ref{1.5}) take on a somewhat classical air.

We should also say that we certainly believe that these results hold for
{\em arbitrary} finite groups $G.$ At the moment we are unable to establish
the general case, but the reader may see that at the cost of more
technical detail but with no further new ideas, Theorem \ref{t1} can be
established for any {\em solvable} group. But to keep the main ideas as
clear as possible we limit our discussion to nilpotent groups,
which may be considered as the first broad class of groups beyond the
abelian groups.

In addition we have the following result which we certainly cannot prove
as yet even for the general solvable group!

\begin{th1}\label{t2} Let $V$ be a simple VOA and  let  $G$ be a faithful,
finite nilpotent group of automorphisms of $V.$ Then there is a Galois
correspondence between subgroups of $G$ and sub VOAs of $V$ which contain
$V^G$ given by the map $H\mapsto V^H.$
\end{th1}

This result was established in [DM2] for $G$ abelian (and $G$
dihedral). We expect that some sort of duality relating the various
terms of (\ref{1.2}) ought to relate Theorems \ref{t1} and \ref{t2}
more closely. We also pointed out in [DM2] that the theory of certain
Von Neumann algebras, whose relation to VOA theory has been remarked on
before (e.g., [MS]), also possesses a Galois theory (cf. [J]) and
references therein). Our earlier comments on commuting algebras
suggest that there may well be a close analogy between these two
theories, at least with regard to the sort of questions we are studying
here.

The proofs of both theorems proceed by induction on the order of $G,$
so that one considers VOAs of the form $V^K$ for various subgroups
$K\subseteqq G.$ It is a result of [DM2] that if $V$ is a simple VOA
then so too is $V^K,$ so that this assumption works well in an
inductive setting. On the other hand, it is no longer appropriate to
assume the unicity of simple $g$-twisted $V$-modules, and we therefore merely
assume that there is a family of twisted sectors which behave in the
`correct' way (cf. Hypothesis \ref{h3.3}).
In this generality, our main result (Theorem \ref{t6.1}), will also apply
to rational VOAs as well as holomorphic VOAs.

The paper proceeds as follows: in Section 2, we review some basic facts from
VOA theory, in particular so called {\em duality,} which plays an
important role.  Section 3 explains the role of Hochschild cohomology,
following [M]. In Section 4 we describe the action of
$D_{\a}(G)$ on $V^*.$ In Section 5 we present the Zhu algebra $A(V)$
together with some variations which are adapted
to orbifold theory. These ideas are developed in [DLM].
Sections 6 and 7 provide the proofs of the two theorems.

\section{Vertex operator algebras and modules}
\setcounter{equation}{0}
In this section we recall the definitions of vertex operator algebras
and modules (cf. [B], [FLM], [DM2]). We also
discuss duality for twisted modules.

A {\it vertex operator algebra} (or VOA) (cf. [B], [FLM]) is a ${\Z}$-graded
vector space
$V=\coprod_{n\in{\Z}}V_n$ such that $\dim\,V_n<\infty$ and $V_n=0$ if $n$ is
sufficiently small, equipped with a linear map
\begin{equation}\label{2.1}
\begin{array}{l}
V \to (\mbox{End}\,V)[[z,z^{-1}]]\\
v\mapsto Y(v,z)=\displaystyle{\sum_{n\in{\Z}}v_nz^{-n-1}}\ \ \ \  (v_n\in
\mbox{End}\,V)
\end{array}
\end{equation}
and with two distinguished vectors ${\bf 1}\in V_0,$ $\omega\in V_2$
satisfying the following conditions for $u, v \in V$:
\begin{eqnarray}
& &u_nv=0\ \ \ \ \ \mbox{for}\ \  n\ \ \mbox{sufficiently large};\label{e2.2}\\
& &Y({\bf 1},z)=1;\label{e2.3}\\
& &Y(v,z){\bf 1}\in V[[z]]\ \ \ \mbox{and}\ \ \ \lim_{z\to
0}Y(v,z){\bf 1}=v;
\end{eqnarray}
\begin{equation}\label{jac}
\begin{array}{c}
\displaystyle{z^{-1}_0\delta\left(\frac{z_1-z_2}{z_0}\right)
Y(u,z_1)Y(v,z_2)-z^{-1}_0\delta\left(\frac{z_2-z_1}{-z_0}\right)
Y(v,z_2)Y(u,z_1)}\\
\displaystyle{=z_2^{-1}\delta
\left(\frac{z_1-z_0}{z_2}\right)
Y(Y(u,z_0)v,z_2)}
\end{array}
\end{equation}
(Jacobi identity) where $\delta(z)=\sum_{n\in {\Z}}z^n$ is
the algebraic formulation of the $\delta$-function at 1, and all binomial
expressions are to be expanded in nonnegative
integral powers of the second variable;
\begin{equation}\label{e2.6}
[L(m),L(n)]=(m-n)L(m+n)+\frac{1}{12}(m^3-m)\delta_{m+n,0}(\mbox{rank}\,V)
\end{equation}
for $m, n\in {\Z},$ where
\begin{equation}
L(n)=\omega_{n+1}\ \ \ \mbox{for}\ \ \ n\in{\Z}, \ \ \
\mbox{i.e.},\ \ \ Y(\omega,z)=\sum_{n\in{\Z}}L(n)z^{-n-2}
\end{equation}
and
\begin{eqnarray}
& &\mbox{rank}\,V\in {\Q};\\
& &L(0)v=nv=(\mbox{wt}\,v)v \ \ \ \mbox{for}\ \ \ v\in V_n\
(n\in{\Z}); \label{3.40}\\
& &\frac{d}{dz}Y(v,z)=Y(L(-1)v,z).\label{3.41}
\end{eqnarray}
This completes the definition. Note that (\ref{e2.6}) says that the
operators $L(n)$ generate a copy of the Virasoro algebra, represented on $V$
with central charge rank\,$V.$
We denote the vertex
operator algebra just defined by $(V,Y,{\bf 1},\omega)$
(or briefly, by $V$). The series $Y(v,z)$ are called {\it vertex operators.}

Let $(V,Y,{\bf
1},\omega)$ are vertex operator algebra. An {\it automorphism} of $V$
is a linear map $g$: $V\to V$  satisfying
\begin{equation}
gY(v,z)g^{-1}=Y(gv,z),\ \ v\in V
\end{equation}
\begin{equation}
g{\bf 1}={\bf 1},\ \ g{\omega}=\omega.
\end{equation}
Let $\Aut(V)$ denote the group of all automorphisms of $V.$

Now each $g$ commutes with the component operators $L(n)$ of $\o,$ and
in particular $g$ preserves the homogeneous spaces $V_n$ which are the
eigenspaces for $L(0).$ So each $V_n$ is a representation module for
$\Aut(V).$

Let $g$ be an  automorphism of the VOA $V$ of order $N.$
Following [FFR] and [D2],
a {\em weak} $g$-{\it twisted}
$module$ $M$ for $V$ is a $\C$-graded
vector space $M=\coprod_{n\in{\C}}M_n$ such that for $\lambda\in\C,$
$M_{n+\l}=0$ for $n\in\frac{1}{N}\Z$
is sufficiently small. Moreover there is a linear map
\begin{equation}
\begin{array}{l}
V\to (\mbox{End}\,M)[[z^{1/N},z^{-1/N}]]\label{map}\\
v\mapsto\displaystyle{ Y_g(v,z)=\sum_{n\in{\frac{1}{N}\Z}}v_nz^{-n-1}\ \ \
(v_n\in
\mbox{End}\,M)}
\end{array}
\end{equation}
satisfying  axioms analogous to (\ref{e2.2})-(\ref{e2.3}) and
(\ref{jac})-(\ref{3.41}). To describe these, let $\eta=e^{2\pi i/N}$ and
set $V^j=\{v\in V|gv=\eta^jv\},$ $0\leq j\leq N-1.$ Thus we have a direct
sum decomposition
\begin{equation}\label{dec}
V=\coprod_{j\in \Z/N\Z}V^j.
\end{equation}
Then we require that for $u,v\in V,$ $w\in M,$
\begin{eqnarray}
& &Y_g(v,z)=\sum_{n\in j/N+\Z}v_nz^{-n-1}\ \ \ \ {\rm for}\ \ v\in V^j;
\label{1/2}\\
& &u_nw=0\ \ \ \mbox{for}\ \ \ n\ \ \ \mbox{sufficiently\ large};\\
& &Y_g({\bf 1},z)=1;
\end{eqnarray}
\begin{equation}\label{jacm}
\begin{array}{c}
\displaystyle{z^{-1}_0\delta\left(\frac{z_1-z_2}{z_0}\right)
Y_g(u,z_1)Y_g(v,z_2)-z^{-1}_0\delta\left(\frac{z_2-z_1}{-z_0}\right)
Y_g(v,z_2)Y_g(u,z_1)}\\
\displaystyle{=z_2^{-1}\left(\frac{z_1-z_0}{z_2}\right)^{-k/N}\delta\left(\frac{z_1-z_0}{z_2}\right)
Y_g(Y(u,z_0)v,z_2)}
\end{array}
\end{equation}
for $u\in V^k.$
Finally, (\ref{e2.6})-(\ref{3.41}) go over unchanged except that in
(\ref{3.40}) we replace $v$ by $w\in M.$ This completes the
definition. We denote this module by $(M,Y_g),$ or briefly by $M.$

\begin{rem}{\rm A $g$-{\em twisted} $V$-module is a weak
$g$-twisted module such that
each homogeneous subspace $M_n$ is finite-dimensional.
A $g$-twisted $V$-module
is a $V$-module if $g=1.$ Moreover,
a $g$-twisted $V$-module restricts to an ordinary $V^0$-module.}
\end{rem}

\begin{rem}\label{r2.2}
{\rm For a weak $g$-twisted $V$-module $M$ and $\lambda\in C,$
$M(\lambda)=\sum_{n\in \frac{1}{N}\Z}M_{\lambda+n}$ is a weak $g$-twisted
submodule of $V$ by (\ref{1/2}). It is clear that $M(\lambda)=M(\mu)$
if and only if $\lambda-\mu\in\frac{1}{N}\Z.$ Moreover,
$M=\coprod_{\lambda\in \C/\frac{1}{N}\Z}M(\l).$ Thus it is enough
to study weak $g$-twisted module of type
$$M=\coprod_{n\in\frac{1}{N}\Z, n\geq 0}M_{c+n}$$
where $c\in\C$ is a fixed and $M_c\ne 0.$ We call $M_c$ the {\em top
level} of $M.$  Note that $M$ has this restricted gradation if
$M$ is irreducible. The same comments also apply to $g$-twisted modules
and ordinary modules.}
\end{rem}

Next we shall present the duality properties for twisted modules.
Let $V$ be a VOA, $g\in \Aut(V)$ of order $N$ and
$W=\coprod_{n\geq n_0}W_n$ a $g$-twisted $V$-module.
Let $W_n^*$ be the dual space of $W_n$ and
$W'=\coprod_{m\geq m_0}W_m^*$ the restricted dual of $W.$ We denote by
$\langle \cdot,\cdot\rangle: W'\times W\to \C$
the restricted paring such that
$\langle W_n^*,W_m\rangle=0$ unless $m=n.$

Set
$$\C[(az_1+bz_2)^{1/N},(az_1+bz_2)^{-1/N}|a,b\in\C, ab\ne 0 ].$$
For each of the two orderings $(i_1,i_2)$ of the set $\{1,2\}$ there
is an injective ring map
$$\iota_{i_1i_2}: R\to \C[[z_1^{1/N},z_1^{-1/N},z_2^{1/N},z_2^{-1/N}]]$$
by which an element $(az_1+bz_2)^{n}\in R$ for $n\in \frac{1}{N}\Z$
is  expanded in nonnegative integral powers of $z_{i_2}.$ Using
the proof of the ``duality'' for a generalized  vertex algebra given
in Chapter 9 (Propositions 9.12 and 9.13) of [DLe] we have
\begin{prop}\label{p2.1} (i) {\bf Rationality}: Let $u\in V^r, v\in V^s
w\in W, w'\in W'$ with $0\leq r,s<N$ and $r,s\in \Z.$
Then there is $f\in R$ of the form
$f(z_1,z_2)=h(z_1,z_2)/z_1^{r/N+m}z_2^{s/N+n}(z_1-z_2)^t$ with
$h(z_1,z_2)\in \C[z_1,z_2]$ and $m,n,t\in \Z$ nonnegative such that
$t$ only depends on $u$ and $v$ and that
\begin{equation}\label{3.1}
\langle w',Y(u,z_1)Y(v,z_2)w\rangle=\iota_{12}f(z_1,z_2).
\end{equation}

(ii) {\bf Commutativity}: If $u,v,w,w',f$ are as in (i) then we also
have
\begin{equation}\label{3.2}
\langle w',Y(v,z_2)Y(u,z_1)w\rangle =\iota_{21}f(z_1,z_2).
\end{equation}

(iii) {\bf Associativity}: In the same notation,
\begin{equation}\label{3.3}
\langle w',Y(Y(u,z_0)v,z_2)w\rangle=\iota_{20}f(z_0+z_2,z_2).
\end{equation}
\end{prop}

Using rationality (i) and associativity (iii) and following the proof
of Proposition 4.1 of [DM2] we have (see also Lemma 5.5 of [DLi]):
\begin{prop}\label{pa}
Let $V$ be a vertex operator algebra and $g$ an automorphism
of order $N.$ Assume that
$M$ is a weak $g$-twisted module generated by $S\subset M$
in the sense that $M$ is the linear span of
$$v_{n_1}^1\cdots v_{n_k}^ks$$
for $v^i\in V,$ $n_i\in\frac{1}{N}\Z,$ $s\in S$ and $k\geq 0.$ Then $M$ is
the linear span of the subset $\{u_ns|u\in V, n\in\frac{1}{N}\Z,s\in S\}.$ In
particular, if $M$ is simple then for each $0\ne w\in M,$
$M$ is spanned as weak $g$-twisted $V$-module by $u_nv$ for
$u\in V,n\in\frac{1}{N}\Z.$ \ \qed
\end{prop}

\section{Group cohomology and VOAs}
\setcounter{equation}{0}
Let $\Irr(V)$ be the set of (isomorphism classes of) inequivalent simple
$V$-modules. Following Section 2 of [DM1], there is a {\em right} action
of $G$ on $\Irr(V)$ (we are taking $G$ to be a subgroup of $\Aut(V)$) given
as follows: If $M=(M,Y_M)\in \Irr(V)$ and $g\in G$ then
\begin{equation}\label{3.1'}
(M,Y_M)\circ g=(M\circ g,Y_M\circ g)
\end{equation}
where by definition $M\circ g=M$ and
\begin{equation}\label{3.2'}
Y_M\circ g(v,z)=Y_M(gv,z).
\end{equation}
More generally, let $\Irr_g(V)$ be the set of inequivalent simple $g$-twisted
$V$-modules, $g\in G.$ Then (\ref{3.1'}) and (\ref{3.2'}) define maps, for
$h\in G,$
$$\Irr_g(V)\times \{h\}\to \Irr_{h^{-1}gh}(V).$$
Thus if ${\cal M}$ is the union of all $\Irr_g(V),$ $g\in G,$ we have a right
action ${\cal M}\times G\to {\cal M}$ of $G$ on ${\cal M}.$ Of course
these definitions hold true for {\em any} $g$-twisted $V$-module: there is a
right $G$-action which preserves the set of simple objects.

Next let $V$ be a VOA with $g\in \Aut(V),$ and let $M=(M,Y_M)$ be a
$g$-twisted $V$-module. An {\em automorphism} of $M$ consists of a
pair $(x,\a(x))$ satisfying the following: $x: M\to M$ and $\a(x):
V\to V$ are linear isomorphisms such that
\begin{equation}\label{3.3'}
\begin{array}{c}
xY_M(v,z)x^{-1}=Y_M(\alpha(x)v,z)\\
\alpha(x)g=g\alpha(x), \a(x){\bf 1}={\bf 1},
\a(x)\omega=\omega
\end{array}
\end{equation}
for $v\in V.$ If $V$ and $M$ are both simple it is easy to see from the
axioms that the following hold: $x\to\a(x)$ is a group homomorphism
$\a$ with kernel consisting of all scalar operators on $M.$ Moreover the
image of $\alpha$ is a group of automorphisms of $V$ that commutes with $g.$
Thus $x$ is then just an isomorphism from $(M,Y_M)$ to $(M,Y_M)\circ\a(x).$

This approach is basically the opposite of that in [DM1] for constructing
projective representations on twisted sectors. We quickly recall the details.

Let $(M,Y_M)\in \Irr_g(V)$ and let $H$ be the subgroup of $\Aut(V)$
which commutes with $g$ and which satisfies $(M,Y_M)\circ h\cong
(M,Y_M),$ all $h\in H.$ This means that there is a linear isomorphism $\phi(h):
M\to
M$ satisfying
\begin{equation}\label{3.4'}
\phi(h)Y_M(v,z)\phi(h)^{-1}=Y_M\circ h(v,z)=Y_M(hv,z)
\end{equation}
for $v\in V.$ Thus $(\phi(h),h)$ is an automorphism of $M,$ and as
explained in Section 2 of [DM1], the simplicity of $M$ together with
Schur's lemma shows that $h\mapsto \phi(h)$ is a projective
representation of $H.$ In effect, $\phi$ is a section of the group
homomorphism $\a.$

In general, given $H$ and $\phi$ above, we denote by $\hat H$ the central
extension of $H$ obtained from $\phi.$ When necessary we let $\a_g$ (not to be
confused with $\a!$) be the corresponding 2-cocycle in $C^2(H,\C^{\times}).$

We next consider the action of $g\in\Aut(V)$ of finite order on
$\Irr_g(V).$ We  will show that it is trivial.
\begin{lem}\label{l3.1} Let $(M,Y_M)\in\Irr_g(V).$ Then $(M,Y_M)\circ g\cong
(M,Y_M).$
\end{lem}

\pf Let $M_c$ be the top level of $M.$ Thus $M_c\ne 0$ while $M_{c+n}=0$
for $n<0.$ Then from Remark \ref{r2.2} we have
$$M=\coprod_{n=0}^{\infty}M_{c+\frac{n}{N}}$$
where $N$ is the order of $g.$ Define $\phi(g): M\to M$ as follows:
\begin{equation}\label{3.5'}
\phi(g)|_{M_{c+\frac{n}{N}}}=e^{-2\pi in/N}.
\end{equation}
{}From (\ref{3.4'}) we see that $(\phi(g),g)$ is an automorphism of $(M,Y_M).$
The lemma follows. \qed

We continue with a simple VOA $V,$ $g\in \Aut(V)$ of finite order $N,$ and
$(M,Y_M)\in \Irr_g(V).$ Let $H$ be as above.
%$$H=\{h\in\Aut(V)|gh=hg, (M,Y_M)\circ h\cong (M,Y_M)\}.$$
Thus $H$ is a group which contains $g.$ Let $h\to \phi(h)$ be the projective
representation of $H$ given by (\ref{3.4'}) and let $\hat H$ be the central
extension of $H$ (by $C^{\times}$) which acts on $M.$

We let $\phi(g)$ be the map (\ref{3.5'}), regarded as an element of $\hat H.$
So $(\phi(g),g)$ is an automorphism of $M.$

\begin{lem}\label{l3.2} $\phi(g)$ lies in the center of $\hat H.$
\end{lem}

\pf Let $\C^{\times}\cong X\leq Z(\hat H)$ be such that $\hat H$ is the central
extension $1\to X\to \hat H\to H\to 1.$ Since $g\in Z(H)$ then clearly we have
$X\<\phi(g)\>\trianglelefteq \hat H$ and $[\phi(g),\hat H]\leq X.$

On the other hand on the top level $M_c$ of $M,$ $\<\phi(g)\>$ is the subgroup
of $X\<\phi(g)\>$ acting trivially. Then $\<\phi(g)\>\trianglelefteq \hat H,$
so that $[\phi(g),\hat H]\leq X\cap \<\phi(g)\>=1.$
The lemma follows. \qed

%There is an action $\rho$ of $\R$ on $M$ defined via
%\begin{equation}\label{3.6'}
%\phi(t)|_{M_{c+\frac{n}{N}}}=e^{-2\pi nt/N}
%\end{equation}
%for $t\in\R.$ This commutes with the action of $\hat H,$ and of
%course we have $\phi(1)=\phi(g).$ Thus there is a push-out diagram
%\begin{equation}\label{3.7'}
%\begin{array}{ccc}
%1 & \mapsto & \phi(g)\\
%\Z & \longrightarrow & \hat H\\
%\downarrow &  & \downarrow \\
%\R & \longrightarrow & \tilde H
%\end{array}
%\end{equation}
%that is $\tilde H=\R\times \hat H/\<(1,\phi(g)^{-1})\>.$ Evidently $\tilde H$
%%%is a central extension of $\C^{\times}$ by a push-out
%\begin{equation}\label{3.8'}
%\begin{array}{ccc}
%1 & \mapsto & g\\
%\Z & \longrightarrow & H\\
%\downarrow &  & \downarrow \\
%\R & \longrightarrow & \bar H
%\end{array}
%\end{equation}

The following axiomatizes one of the basic situation with which we are
concerned.
\begin{hy}\label{h3.3}
 $V$ is a simple VOA and $G\leq \Aut(V)$ a finite group of
automorphisms. For each $g\in G$ there is $(M(g),Y_g)\in\Irr_g(V)$ such that,
for all $h\in G,$
\begin{equation}\label{3.9'}
(M(g),Y_g)\circ h\cong (M(h^{-1}gh),Y_{h^{-1}gh}).
\end{equation}
\end{hy}

\begin{rem}
It is a well-known conjecture that (\ref{3.9'}) always holds if $V$ is
{\em holomorphic}, $M(g)$ being the {\em unique} element of  $\Irr_g(V)$ in
that case.
\end{rem}

%In the presence of Hypothesis \ref{h3.3}, the point of Lemma \ref{l3.2} is
%this: the group $H$ above may be taken to be the {\em centralizer}
%$C_G(g)$ of $g$ in $G.$ Then (\ref{3.8'}) defines the so-called {\em
%extended centralizer} of $g$ (cf. Section 2.15 of [B?] or
%[L]).Together with (\ref{3.7'}), we see that $\tilde H$ is a central
%extension of the extended centralizer $\tilde C_G(g).$ The upshot is
%that from Hypothesis \ref{h3.3} we can deduce the existence of an element in
%%t%he direct
%sum
%\begin{equation}\label{3.10'}
%\oplus_{g^*}H^2(\tilde C_G(g),\C^{\times})
%\end{equation}
%where the sum runs over a fixed choice of $g$ in each conjugacy class of $G.$
%%Moreover, the cohomology ``comes from'
%$H^2(C_G(g),\C^{\times})\simeq H^3(C_G(g),\Z).$ In this way, using results of
%Burghele [Bu], Hypothesis \ref{h3.3} determines an element in the third {\em
%%c%yclic} cohomology group $HC^3(\Z G).$

For the rest of the paper we will concentrate on the element $\a$ of
\begin{equation}\label{3.11'}
HH^3(\Z G)\simeq \bigoplus_{*}H^2(C_G(g),\C^{\times})
\end{equation}
(where $HH$ stands for {\em Hochschild} cohomology and the sum in (\ref{3.11'})
runs over one $g$ in each conjugacy class of $G$) determined by
Hypothesis \ref{h3.3}. Specifically, for $g\in G,$ let $\a_g\in
C^2(C_G(g),\C^{\times})$
be the 2-cocycle corresponding to the projective representation $\phi=\phi_g$
(\ref{3.4'}).

More precisely, using (\ref{3.9'}), we have maps $\phi_g(h)$: $M(g)\to
M(hgh^{-1})$ for
$g,h\in G$ and
\begin{equation}\label{3.12'}
\phi_g(h)Y_g(v,z)\phi_g(h)^{-1}=Y_{hgh^{-1}}(hv,z).
\end{equation}
Compatibility yields
\begin{equation}\label{3.13'}
\phi_g(hk)=\a_g(h,k)^{-1}\phi_{kgk^{-1}}(h)\phi_g(h)
\end{equation}
for some $\a_g(h,k)\in \C^{\times}$ (even $S^1$) satisfying
\begin{equation}\label{3.14'}
\a_g(hk,l)\a_{lgl^{-1}}(h,k)=\a_{g}(h,kl)\a_g(k,l).
\end{equation}
If $h,k,l\in C_G(g)$ then (\ref{3.14'}) reduces to the assertion that $\a_g\in
C^2(C_G(g),\C^{\times});$ it is the 2-cocycle associated with $\phi_g.$

\section{The twisted quantum double}
\setcounter{equation}{0}
We continue to assume Hypothesis \ref{h3.3}. Associated to this situation is
the
{\em algebraic} space
\begin{equation}\label{4.1}
V^*=\oplus_{g\in G}M(g).
\end{equation}

We will define a certain associative algebra $D_{\a}(G),$ the {\em twisted
quantum double,} where $\a$ is the element in $HH^3(\Z G)$ discussed in
Section 3. We will then prove
\begin{thm}\label{t4.1}
The following hold:

(i) $V^*$ is a module over $D_{\a}(G).$

(ii) Every simple $D_{\a}(G)$-module occurs as a submodule of $V^*.$

(iii) If $V^G$ is the sub VOA of $G$-invariants of $V$ then $V^*$ is
a $V^G$-module and the actions of $D_{\a}(G)$ and $V^G$ on $V^*$ commute.
\end{thm}

Introduce the complex group algebra $\C[G]$ and its dual $\C[G]^*$ with
basis $e(g),$ $g\in G$ satisfying $e(g)e(h)=e(g)\delta_{g,h}.$ The group
of units of $\C[G]^*$ is the multiplicative group
\begin{equation}\label{4.2}
U=\{\sum_{g\in G}a_ge(g)|a_g\in \C^{\times}\}.
\end{equation}
The group $G$ acts on $\C[G]^*$ on the right via $e(g)\cdot h=e(h^{-1}gh).$
This action preserves $U,$ which thereby becomes a multiplicative right
$\Z G$-module.

Define $\a:$ $G\times G\to U$ via
\begin{equation}\label{4.3}
\alpha(h,k)=\sum_{g\in G}\a_g(h,k)e(g).
\end{equation}
Then (\ref{3.14'}) is equivalent to
\begin{equation}\label{4.4}
\alpha(hk,l)\alpha(h,k)^l=\alpha(h,kl)\alpha(k,l)
\end{equation}
which says that $\a\in C^2(\Z G,U),$ the group of 2-cocycles on $G$ with
values in $U.$ We leave it to the reader for sort-out the relation between
$H^2(G,U)$ and $HH^3(\Z G).$

Following [M], the {\em twisted quantum double} $D_{\a}(G)$ has underlying
space $\C[G]\otimes\C[G]^*$ with multiplication
\begin{equation}\label{4.5}
a\otimes e(x)\cdot b\otimes e(y)=\alpha_y(a,b)ab\otimes e(b^{-1}xb)e(y).
\end{equation}
If $\alpha=1$ this reduces to the usual quantum double [Dr]. $D_{\a}(G)$ is a
semi-simple associative algebra.

If $m\in M(g),$ $g\in G,$ we define
\begin{equation}\label{4.6}
a\otimes e(x)\cdot m=\delta_{x,g}\phi_x(a)m
\end{equation}
where $\phi_g(a): M(g)\to M(aga^{-1})$ is as before. Using (\ref{3.13'}) and
(\ref{4.5}) shows that (\ref{4.6}) defines a left action of $D_{\a}(G)$ on
$V^*.$ This proves (i).

Part (ii) is easy. We have already pointed out in Section 2 that $g$-twisted
modules are ordinary $V^G$-modules. So certainly $V^*$ is a $V^G$-module.

Using (\ref{3.12'}) and (\ref{4.6}) we find that for $v\in V$ we have
\begin{equation}\label{4.7}
a\otimes e(x)\bar Y(v,z)=\bar Y(av,z)a\otimes e(x)
\end{equation}
as operators on $V^*,$ where we have used $\bar Y(v,z)$ to denote the operator
on $V^*$ which acts on $M(g)$ as $Y_g(v,z).$ In particular, it is clear that if
$v\in V^G$ then $a\otimes e(x)$ commutes with $\bar Y(v,z).$ So (iii) of the
theorem holds.

We turn our attention to (ii). Fix $g\in G$ and set $C=C_G(g).$ If
$a,b\in C$ then from (\ref{4.5}) we see that $D_{\a}(G)$ contains the
$\alpha_g$-{\em twisted group algebra} $\C_{\alpha_g}[G]$ which has
basis indexed by $a\in C$ and multiplication $a\cdot b=\a_g(a,b)ab.$
More precisely, if $S(g)$ is the subspace of $D_{\a}(G)$ with basis
$a\otimes e(g)$ for $a\in C,$ then $S(g)$ is isomorphic to $\C_{\a_g}[C]$
via $a\otimes e(y)\mapsto a.$

Now (\ref{4.6}) defines a left action of $S(g)$ on $M(g).$ If we also let
$D(g)$ be the subspace of $D_{\a}(G)$ spanned by $a\otimes e(g)$ for all
$a\in G$ then $D(g)$ is an $D_{\a}(G)-S(g)$-bimodule.

\begin{lem}\label{l4.2}
Let $K$ be the conjugacy class of $G$ that contains $g.$ Then there is an
isomorphism of $D_{\a}[G]$-modules:
$$D(g)\otimes_{S(g)}M(g)\simeq \oplus_{h\in K}M(h)$$
given by $a\otimes e(g)\otimes m\to \phi_g(a)m$ for $a\in G.$ \ \qed
\end{lem}

This is an easy calculation. The point is that it is shown in [M] (see
also [DPR]) that the simple $D_{\a}(G)$-modules are precisely the
modules $D(g)\otimes_{S(g)}X$ where $X$ ranges over the simple
$S(g)$-modules and $g$ ranges over one element in each conjugacy class of
$G.$ Thus after Lemma \ref{l4.2}, Theorem \ref{t4.1} (ii) is a consequence
of
\begin{lem}\label{lem4.3} Every simple $S(g)$-module is a submodule of $M(g).$
\end{lem}

We will need
\begin{lem}\label{l4.4} For $m\geq 1$ let $u_1,...,u_m$ and $w_1,...,w_m$
be non-zero vectors in $V$ and linearly independent vectors in $M(g)$
respectively. Then
$$\sum_{i=1}^mY_g(u_i,z)v_i\ne 0.$$
\end{lem}

\pf This is the same as the proof of Lemma 3.1 of [DM2]. We just have to
replace the duality arguments for simple $V$-modules used in [DM2] by
the corresponding duality statements for twisted modules as stated in
Section 2.  \qed

{\em Proof of Lemma \ref{4.3}:} We use Theorem 2 of [DM2] which tells us that
every simple $\C[C]$-module is a submodule of $V.$ Let $S=S(g).$

Let $A$ be a simple $S$-module which is a submodule of $M(g)$ and let $B$ be
any simple $S$-module. We can find a simple $\C[C]$-module
$D$ such that $\Hom_S(A^*\otimes B,D)\ne 0.$ Thus $B$ is a submodule of the
$S$-module $D\otimes A.$

We may take $D\subset V,$ so that expressions of the form
$\sum_{i=1}^mY_g(v_i,z)w_i$ with $v_i\in D$ and $w_i\in A$ can be used. We can
thus complete the
proof using Lemma \ref{l4.4} as in the proof of Theorem 2 of [DM2]. \qed

\section{Zhu algebras}
\setcounter{equation}{0}

In [Z], Zhu introduced an associative algebra $A(V)$ associated to a VOA $V$
which is extremely useful in studying the representation theory of $V.$
There are analogues for $g$-twisted modules which are developed in [DLM],
which we review here.

Fix a VOA $V$ and an automorphism $g$ of $V$ (of finite order). For $u,v\in V$
with $u$ homogeneous, define a product $u*v$ as follows:
\begin{equation}\label{5.1}
u*v={\rm Res}_zY(u,z)\frac{(z+1)^{{\rm wt}\,u}}{z}v
 =\sum_{i=0}^{\infty}{\wt\,u\choose i}u_{i-1}v.
\end{equation}
Then extends (\ref{5.1}) to a linear product $*$ on $V.$

For $0<r\leq 1$ let $V^r$ the eigenspace of $g$ with eigenvalue
$e^{2\pi ir},$ that is,
\begin{equation}\label{5.2}
V^r=\{v\in V|gv=e^{2\pi ir}u\}.
\end{equation}
Define a subspace $O_g(V)$ of $V$ to be the linear span of all elements
$u\circ_gv$ of the following type if $u$ is homogeneous, $u\in V^r$
and $v\in V:$
\begin{equation}\label{5.3}
u\circ_gv=\left\{
\begin{array}{ll}
{\rm Res}_zY(u,z)\frac{(z+1)^{{\rm wt}\,u}}{z^2}v, & {\rm if}\ r=1\\
{\rm Res}_zY(u,z)\frac{(z+1)^{\wt\,u+r-1}}{z}v, & {\rm if}\ r<1.
\end{array}\right.
\end{equation}
Note that if $r<1$ then $u\circ_g{\bf 1}=u.$ Thus we have
\begin{lem}\label{l5.0}
If $V^{\<g\>}$ is the sub VOA of $g$-invariants of $V$ then
$V=V^{\<g\>}+O_g(V).$ \qed
\end{lem}

Let $(M,Y_g)$ be a weak $g$-twisted $V$-module. For homogeneous $u\in V,$
the component operator $u_{\wt\,u-1}$ preserves each homogeneous subspace
of $M$ and in particular acts on the top level $M_c$ of $M$ (cf. Section 2).
Let $o_g(u)$ be the restriction of $u_{\wt\,u-1}$ to $M_c,$ so that
we have a linear map
\begin{equation}\label{5.4}
\begin{array}{lll}
V &\to & \End(M_c)\\
u&\mapsto &o_g(u).
\end{array}
\end{equation}
Note that if $u\in V^r$ with $r<1$ then $o_g(u)=0$ from (\ref{1/2}).

Set $A_g(V)=V/O_g(V).$ Then we have [DLM]:
\begin{thm}\label{t5.1} (i) $A_g(V)$ is an associative algebra with
multiplication induced from
$*$ and the centralizer $C(g)$ of $g$ in $\Aut(V)$ induces
a group of algebra automorphisms of $A_g(V).$

(ii) $u\mapsto o_g(u)$ gives a representation of $A_g(V)$ on $M_c.$
Moreover, if every weak $g$-twisted module is completely reducible,
$A_g(V)$ is semisimple.

(iii) $M\mapsto M_c$ gives a bijection between the set of equivalence
classes of simple weak $g$-twisted $V$-modules and the
 set of equivalence
classes of simple $A_g(V)$-modules. \qed
\end{thm}

 Next, with $V$ as before and with $G$ an automorphism group of $V,$ we define
\begin{equation}\label{5.5}
O_G(V)=\cap_{g\in G}O_g(V),\ \
A_G(V)=V/O_G(V).
\end{equation}
\begin{lem}\label{l5.2}
$A_G(V)$ is an associative algebra with respect to product $*.$
\end{lem}

\pf From Theorem \ref{t5.1} (i), $O_g(V)$ is a 2-sided ideal of $V$ with
respect to $*.$ Moreover if $u,v,w\in V$ then
$(u*v)*w-u*(v*w)\in O_g(V)$ for all $g\in G.$ The lemma follows.
\qed

{}From Theorem \ref{t5.1} we see that the simple modules for $A_G(V)$
correspond
to the top levels of weak simple $g$-twisted $V$-modules for {\em all} $g\in
G.$

We will need to consider the algebra $A_G(V),$ in a special case, in the
course of the proof of Theorem \ref{t6.1}.
\begin{thm}\label{t5.3} Let $V$ be a VOA and $G$ a finite group of
automorphisms
of $V$ with $g\in G$ in the center of $G.$ Let $M,N$ be two simple $g$-twisted
$V$-modules with $X,Y$ the top levels of $M$ and $N$ respectively. Assume that
the weight of $Y$ is less than or equal to the weight of $X.$ Let
$N'$ be the $V^G$-submodule of $N$ generated by $Y.$ Exactly one of the
following holds:

(i) $M\circ h\simeq N$ (isomorphism of $g$-twisted $V$-modules) for some
$h\in G.$

(ii) $\Hom_{V^G}(N',M)=0.$
\end{thm}

\pf As $g\in Z(G)$ then $G$ induces algebra automorphisms of $A_g(V).$ Assume
that (i) fails. So $N$ is not isomorphic to $M\circ h$ for any $h\in G.$

Let $M=M_1,...,M_r,$ $N=N_1,...,N_s$ be the distinct conjugates of $M$ and
$N$ under the action of $G,$ and let $X=X_1,...,X_r$, $Y=Y_1,...,Y_s$ be the
corresponding top levels. So these $r+s$ spaces afford inequivalent simple
$A_g(V)$-modules (Theorem \ref{t5.1}), and on their direct sum $A_g(V)$
realizes the algebra $\oplus_{i=1}^r\End(X_i)\oplus
\oplus_{j=1}^s\End(Y_j).$

One computes that the algebra $C$ of $G$-invariants on $
\oplus_{i=1}^r\End(X_i)$ is non-zero and satisfies $\ann_C(Y)=C,$
$\ann_X(C)=0.$
Note that the image of the natural map $A(V^G)\to A_g(V)$ contains $C.$
We conclude that $\Hom_{A(V^G)}(Y,X)=0.$

But if (ii) of the Theorem fails then there is $f\in \Hom_{V^G}(N',M)$
satisfying $0\neq f(N')\subset M.$ As $Y$ generates $N'$ then $f(Y)\ne 0,$ and
by the assumption that the weight of $Y$ is less than or equal to weight of $X$
we conclude that $0\ne f(Y)\subset X.$ Thus $f$ induces a nonzero element of
$\Hom_{A(V^G)}(Y,X),$ which is the desired contradiction. \qed

\section{Proof of Theorem 1}
\setcounter{equation}{0}
We continue with the assumptions and notation of Hypothesis \ref{h3.3}. As
described in Section 3, there is a projective representation of
$C_G(g)$ on $M(g)$ for $g\in G,$ corresponding to a 2-cocycle $\a_g\in
C^2(C_G(g),S^1).$ In this way $M(g)$ becomes a module over the twisted
group algebra $\C_{\alpha_g}[C_G(g)].$ Moreover this algebra commutes with
the vertex operators of $V^G.$

\begin{thm}\label{t6.1} Assume that $G$ is nilpotent. Then the following hold:

(A) There is a decomposition of $M(g)$ into simple modules over
$\C_{\a_g}[C_G(g)]\otimes V^G$ of the form
\begin{equation}\label{6.1}
M(g)\simeq \oplus_{\chi_g}M_{\chi_g}\otimes V_{\chi_g}
\end{equation}
where $\chi_g$ ranges over the simple characters of $\C_{\a_g}[C_G(g)]$,
$M_{\chi_g}$ is a module over $\C_{\chi_g}[C_G(g)]$ which affords $\chi_g,$
and $V_{\chi_g}$ is a simple $V^G$-module.

(B) Let $g,h\in G.$ Then there is an isomorphism of $V^G$-modules
$V_{\chi_g}\simeq V_{\psi_h}$ if and only if there is $k\in G$ such that
$k(\chi_g)=\psi_{h}.$
\end{thm}

\begin{rem} The notation $k(\chi_g)=\psi_h$ means that $h=kgk^{-1}$ and also
if $z\in C_G(g)$ then $\chi_g(z)=\psi_h(kzk^{-1}).$
\end{rem}

Let us explain how Theorem \ref{t6.1} implies Theorem \ref{t1}. As explained
in Section 4, if we fix a choice of $g$ in each conjugacy class of $G$ then
we have
\begin{equation}\label{6.2}
V^*=\oplus_{g}\Ind_{S(g)}^{D(g)}M(g).
\end{equation}
Here, we are using the notation of Section 4 (cf. Lemma \ref{l4.2}) and
(\ref{6.2}) is an isomorphism of $D_{\a}(G)$-modules. Since $D_{\a}(G)$
commutes with $V^G$ on $V^*,$ it follows from
(\ref{6.1}) and results from Section 4 that $V^*$ is a $D_{\a}(G)\otimes
V^G$-module with decomposition into simple modules
\begin{equation}\label{6.3}
V^*=\oplus_{g}(\Ind_{S(g)}^{D(g)}M(g))\otimes V_{\chi_g}.
\end{equation}
Moreover {\em every} simple $D_{\a}(G)$-module occurs in (\ref{6.3}) with
non-zero multiplicity. Thus the map
$$ \Ind_{S(g)}^{D(g)}M(g)\stackrel{\phi}{\to} V_{\chi_g}$$
induces a map
$$\{{\rm simple}\ D_{\a}(G){\rm -modules}\}\stackrel{\phi}{\to}
\{{\rm simple}\ V^G{\rm -modules}\}.$$

Theorem \ref{t6.1} (B) tells us that $\phi$ is an injection, so that
$\phi$ is a bijection from the set of  simple $D_{\a}(G)$-modules
to Im${\phi}$, which is just the set of simple $V^G$-modules contained
in $V^*$ (for by Theorem \ref{t6.1} (A), $V^*$ is completely
reducible as $V^G$-module). Thus we have constructed the bijection $\phi$
of (\ref{1.5}). And again since the
categories ${\cal V}^*(G)$ and Mod-$D_{\a}(G)$
are semi-simple then $\phi$ extends to the equivalence of categories
(\ref{1.3}). So then Theorem \ref{t1} is proved.

Note, in fact, that we have proved Theorem \ref{t1} for any simple
VOA under the assumptions of Hypothesis \ref{h3.3}. We thus expect (cf. Section
2
of [DM1]) that our results will apply to any simple rational VOA $V$ for which
$G$ is a group of {\em inner} automorphism of $V$ (cf. [DM1] for
the definition of inner automorphism).

We record some standard facts about finite nilpotent group (see [G], for
example) which we often use without comment.

\begin{lem}\label{l6.2} Let $G$ be a finite nilpotent group. Then

(i) Every maximal subgroup of $G$ is normal and of prime index in $G.$

(ii) Subgroups and quotient groups of $G$ are nilpotent.

(iii) Central extensions of $G$  are nilpotent.

(iv) If $\chi$ is a simple character of $G$ of degree greater than one, then
there is a maximal subgroup $H$ of $G$ and a simple character $\psi$ on $H$
such that $\chi=\Ind_H^G(\psi).$ \ \qed
\end{lem}

We now begin the proof of Theorem \ref{t6.1}, using induction on
$|G|.$ If $|G|=1$ there is nothing to prove. We frequently use the
following facts: if $H\leq G$ then the invariant sub VOA $V^H$ is itself
simple (Theorem 4.4 of [DM2]); moreover if $H\nor G$ then $G/H$ is a
{\em faithful} group of automorphisms of $V^H$ (Proposition 3.3 of
[DM2]). Thus  the assumptions about the pair $(V,G)$ underlying Theorem
\ref{t6.1} also apply to the pair $(V^H,G/H)$ if $H\nor G.$

We start with the proof of (A). So fix $g\in G$ and let $C=C_G(g),$ $M=M(g).$
We must
establish the decomposition (\ref{6.1}) together with the simplicity of
the $V^G$-module $V_{\chi_g}$ for each $\chi_g.$

{\em Case A1: $C$ is a proper subgroup of $G.$}

In this case let $C<H<G$ with $H$ a
maximal subgroup of $G.$ By Lemma \ref{l6.2} (i) we have $H\nor G$ and $[G:H]$
is a prime $p.$ Since $C=C_{H}(g),$ induction tells us that there is
a decomposition
\begin{equation}\label{6.4}
M=\oplus_{\chi_g}M_{\chi_g}\otimes V_{\chi_g}'
\end{equation}
where $V_{\c_g}'$ is a simple $V^H$-module.

Choose $x\in G\setminus H.$ A simple argument using the containments
$C<H\nor G$ shows that the element $x^{-1}gx$ is not conjugate to $g$ in
$G.$ Since $M\circ x=M(x^{-1}gx)$, it follows from our induction assumption
of (B) that $x$ induces an automorphism of $V^H$ of prime order $p$ such that
$V_{\chi_g}$ and $x(V_{\chi_g})$ are inequivalent simple $V^H$-modules.

Now apply Theorem 6.1 of [DM2] to conclude that the restriction of
$V_{\chi_g}$ from $V^H$ to $V^G$ remains irreducible. Then (\ref{6.4}) is the
desired decomposition of $M.$

So we may now assume that $C=G,$  that is $g\in Z(G).$ For each simple
character $\chi$ of $\C_{\a_g}[G]$ let $M^{\chi}$ denote the $\chi$-homogeneous
component of $M.$ We need to show, then,  that
\begin{equation}\label{6.5}
M^{\chi}=M_{\chi}\otimes V_{\chi}
\end{equation}
for some simple $V^G$-module $V_{\c}.$

{\em Case A2: $\chi$ has degree greater than 1.}

In this case there is
$H\nor G$ with $[G:H]=p,$ a prime, together with a simple character $\psi$
of $\C_{\a_g}[H]$ such that $\chi$ is induced from $\psi.$ Note that we
identify $\alpha_g$ with its restriction to an element of $C^2(H,S^1).$
Note also that we necessarily have $g\in H$ in this situation, as $g$ acts
as a non-zero scalar on any simple $\C_{\a_g}[G]$-module (Lemma \ref{l3.2}).

By induction there is a decomposition of $M$ as $V^H$-module such that, with
earlier notation,
$$M^{\psi}=M_{\psi}\otimes V_{\psi}$$
where naturally $M_{\psi}$ is the appropriate simple $C_{\alpha_g}[H]$-module,
and $V_{\psi}$ is a simple $V^H$-module.

If $x\in G\setminus H$ then we necessarily have that the characters
$x^i(\psi)$ for $i=0,...,p-1$ are pairwise inequivalent simple characters of
$\C_{\a_g}[H]$ (cf. Lemma 6.2 of [DM2]). So again Theorem 6.1 of [DM2]
tells us, since $x(V_{\psi})\simeq V_{x(\psi)}$ as $V^H$-modules, that the
restriction of $V_{\psi}$ to $V^G$ is irreducible. So there are the following
isomorphisms of $V^G$-modules:
\begin{eqnarray*}
& &M^{\chi}=\oplus_{i=0}^{p-1}M^{x^i(\psi)}\simeq
\oplus_{i=0}^{p-1}M_{x^i(\psi)}\otimes V_{x^i(\psi)}\\
& &\hspace {1 cm}\simeq  \oplus_{i=0}^{p-1}M_{x^i(\psi)}\otimes V_{\psi}\\
& &\hspace {1 cm}\simeq M_{\chi}\otimes V_{\chi}.
\end{eqnarray*}
(in the last line we used
$\chi=\Ind_{\C_{\alpha_g}[H]}^{\C_{\alpha_g}[G]}(\psi)=\oplus_{i=0}^{p-1}x^i(\psi),$ and we identified $V_{\chi}$ with the
restriction of $V_{\psi}$ to $V^G.$) So (\ref{6.5}) holds in this case.

{\em Case A3: $\chi$ has degree 1.}

In this case we need to show that $M^{\chi}$ itself is a simple $V^G$-module.
This follows from Proposition \ref{pa} and the proof of Theorem 4.4
of [DM2].

This completes the inductive proof of (A), and it remains to prove (B). So
let $g,h\in G$ be such that the two $V^G$-modules $V_{\chi_g}$ and
$V_{\psi_h}$ are isomorphic. We must show that $k(\chi_g)=\psi_h$ for some
$k\in G.$ Let $C=C_G(g),$ $D=C_G(h),$ $M=M(g)$ and $N=N(h).$ Note that
$V_{\cg}$ and $V_{\ph}$ have top levels of the same weight.

{\em Case B1: $\<C,D\>\ne G.$}

As before there is $H\nor G$ with $[G:H]$ a prime $p$ and $\<C,D\><H.$ Let
$x\in G\setminus H.$

If $V_{\cg}$ is $V^H$-isomorphic to $x^i(V_{\ph})$ for some $i,$ then by
induction $k(\chi_g)=x^i(\psi_h)$ for some $k\in H$ and we are done. So we may
assume that $V_{\cg}$ is {\em not} $V^H$-isomorphic to $x^i(V_{\ph})$ for
any $i.$

Now apply Theorem \ref{t5.3} to the action of $G/H$ on the VOA $V^H$ to
conclude that
$\Hom_{V^G}(V'_{\cg},V_{\ph})=0$ where $V_{\cg}'$ is the $V^G$-submodule of
$V_{\cg}$ spanned by the top level. As $V_{\cg}\simeq V_{\ph}$ as $V^G$-modules
this is the desired contradiction.

{\em Case B2: $\<h,C\>\ne G.$}

In this case $g$ and $h$ cannot be conjugate in $G,$
so certainly there is no $k\in G$ with $k(\chi_g)=\psi_h.$ Now proceed
as in Case B1.

{\em Case B3: $C\ne G.$}

After case B2 we may assume that $C<H\nor G$ with
$[G:H]=p,$ a prime, and $h\not\in H.$ In this case we get the usual
decomposition (\ref{6.1}) of $M$ into $V^H$-modules for simple
$V^H$-modules $V_{\chi_h}.$ But the analogous decomposition does {\em
not} hold for $N,$ since $h\not\in H.$ There is a decomposition of $N$ into
$V^G$-modules, by part (A), but as a $V^H$-module the summands of $N$
become $\bar h$-twisted $V^H$-modules, where $\bar h$ is the automorphism
of $V^H$ (of order $p$) induced by $h.$

If $D_0=D\cap H$ then $D=D_0\<h\>,$ and since $h$ lies in the center
of $D$ then the restriction of every irreducible character of $D$ to
$D_0$ remains irreducible. So for an irreducible character $\a$ of
$D_0,$ $\Ind_{D_0}^D(\a)=\sum_{i=0}^{p-1}\chi \lambda^i$ where $\chi$
is some irreducible character of $D$ contained in $\Ind_{D_0}^D(\a),$
and $\lambda$ generates the group of characters of $D/D_0$ ($\simeq
\Z_p$). From this we can see that $N$ decompose into $\bar h$-twisted
$V^H$-modules as follows:
\begin{equation}\label{6.6}
N=\oplus_{\ph}(\oplus_{i=0}^{p-1}M_{\ph}\otimes V_{\ph \lambda^i}).
\end{equation}
Here, $\ph$ ranges over certain simple characters of $\C_{\ah}[D],$
$M_{\ph}$ is a $\C_{\ah}[D]$-module affording $\ph,$ and $\l$ is as before.
(\ref{6.6}) is supposed to mean that $\oplus_{i=0}^{p-1}M_{\ph}\otimes V_{\ph
\lambda^i}$ is an $\bar h$-twisted $V^H$-module; the individual
summands are not.

Let $T$ be the top level of the $V^H$-module $V_{\cg}$ (\ref{6.1})
and $U$ the top level of the simple $\bar h$-twisted $V^H$-module
generated by $V_{\ph}.$ So $T$ and $U$ afford simple modules
for the algebra $A_{G/H}(V^H)$ (cf. (\ref{5.5})).

Now $h$ induces an automorphism of $A_{G/H}(V^H),$ and as in Case A1,
since $h(V_{\cg})\not\simeq V_{\cg}$ as $V^H$-modules then $T$ and
$h(T)$ afford inequivalent $A_{G/H}(V^H)$-modules.  On the other hand
$h(U)\simeq U$ since $h$ leaves invariant each of the summands in
(\ref{6.6}). So $T$ and $U$ necessarily afford inequivalent
$A_{G/H}(V^H)$-modules. Now the argument of Theorem 6.1 of [DM2]
together with Lemma \ref{l5.0} yields that $T$ and $U$ yield inequivalent
$A(V^G)$-modules and we are done as before.

After Case B3 we may assume that both $g$ and $h$ lie in $Z(G),$ i.e.,
$C=D=G.$

{\em Case B4: $\cg$ has degree greater than 1.}

Let $(H,\a)$ be such that
$H\nor G,$ $[G:H]=p,$ a prime and $\a$ is a (simple) character of
$\C_{ag}[H]$ satisfying $\Ind(\alpha)=\cg.$ Note that $g$ lies in the
center of $\C_{\a_g}[G]$ (see Lemma \ref{l3.2}) hence lies in $H.$

If there is $k\in H$ such that $\Ind(k(\a))=\ph$ then we are done. So
we may assume that this is not the case. Now as $V^H$-module we have
(in the notation of (\ref{6.5})) $$M^{\a}=M_{\a}\otimes V_{\a}$$ for a
simple $V^H$-module $V_{\a}.$ Since $g(\a)\ne \a$ for $g\in G\setminus H$
then restriction of $V_{\a}$ to $V^G$ is simple by Theorem 6.1 of
[DM2] once more. Now if $h\in H$ then we are done by another application of
Theorem \ref{t5.3}.

So we may take $h\in G\setminus H.$ In this case the argument
follows the same lines, but we have to use the variation employed  in case B3.
We omit the straightforward details.

{\em Case B5: $\<g,h\>\ne G.$}

After Case B4 we may assume that both $\cg$ and $\ph$
have degree 1. We can complete the proof in this case using arguments of
[DM2]. Namely, choose a subgroup $H$ with $\<g,h\><H$ and $H$ of prime index
$p,$ so that $H\nor G,$ and let
$\sigma,\tau$ be the restriction of $\cg,\ph$ to $\C_{\ag}[H]$
and $\C_{\ah}[H]$ respectively. If we let $\lambda$ generate the character
group of $G/H$ then we have $\Ind(\sigma)=\oplus_{i=0}^{p-1 }\l^i\cg$ and
$\Ind(\tau)=\oplus_{i=0}^{p-1 }\l^i\ph$ (cf. Lemma 6.2 of [DM2]).

Let $V_{\l^i\cg}$ and $V_{\l^i\ph}$ be the appropriate $V^G$-modules
in $M$ and $N$ respectively and let $V_{\sigma}$ and $V_{\tau}$ be the
corresponding $V^H$-modules. So we have
\begin{equation}\label{6.7}
V_{\sigma}=\oplus_{i=0}^{p-1}V_{\l^i\cg},\ \
V_{\tau}=\oplus_{i=0}^{p-1}V_{\l^i\ph}
\end{equation}
as $V^G$-modules.

Suppose that $\phi: V_{\cg}\to V_{\ph}$ is a $V^G$-isomorphism. Choose
$0\ne w\in V_{\cg}$ and consider the $V^H$-submodule of
$V_{\sigma}\oplus V_{\tau}$ generated by $(w,\phi(w)).$ We have
$V^H=\oplus_{i=0}^{p-1}V^{\l^i}$ where $V^{\l^i}$ is the subspace of $V$
transforming according to the character $\l^i,$ and if $u\in V^{\l^i}$ then
any component operator $u_n$ of $u$ satisfies $u_nw\in V_{\l^i\cg}$
and $u_n\phi(w)\in V_{\l^i\ph}.$

Now the argument used in proof of Theorem 5.1 of [DM2] shows first that
$V_{\sigma}\simeq V_{\tau}$ as $V^H$-modules, so that $g=h$ by induction; and
then that $\cg=\ph,$ as required.

This reduces us to the case that $G=\<g,h\>$ is abelian.

{\em Case B6: $\<g\>\ne G.$}

In this case let $g\in H\nor G$ with $[G:H]$ a prime
$p.$ We still have the decompositions (\ref{6.7}), but as far as $V^H$ is
concerned, $V_{\tau}$ is an $\bar h$-twisted $V^H$-module, where
$\bar h$ is the automorphism of $V^H$ induced by $\bar h.$ But in any case we
can still carry out the argument of the last case, leading to the conclusion
that there is a proper subspace of $V_{\sigma}\oplus V_{\tau}$ invariant
under all $u_n$ for $u\in V^H$ and $n\in \Z$ and containing $(w,\phi(w)).$
This is impossible.

{\em Case B7: $\<g\>=G.$}

In this case, since $G$ is cyclic then the 2-cocycles
$\ah$ and $\ah$ may be taken to be trivial. Now the proof of Theorem 5.1
of [DM2] completes the argument.

This finally completes the proof of Theorem \ref{t6.1}.

\section{Proof of Theorem \ref{t2}}
\setcounter{equation}{0}
In [DM2] the authors suggested that there should be a Galois correspondence
for finite groups of automorphisms of simple VOAs and established such a result
for abelian and dihedral group. In lieu of a deeper understanding of this
phenomenon, we will establish in this section that
such a correspondence holds for
nilpotent groups. Namely we will prove Theorem \ref{t2}.

For each integer $n$ there is a linear map
$$\mu_n: V\otimes V\to V$$
defined by $\mu_n(v\otimes w)=v_nw$ for $v,w\in V$ and with $v_n$ the
$n$th component operator of $v$ in $Y(v,z)=\sum_{n\in\Z}v_nz^{-n-1}.$
Moreover the maps $\mu_n$ are $G$-invariant whenever $G$ is a group of
automorphism of $V,$ where as usual $G$ acts on $V\otimes V$ via $g(v\otimes
w)=gv\otimes gw.$

\begin{lem}\label{l7.1}
Suppose that $v,w\in V$ and $v\otimes w$ is not $G$-invariant. Then there
is $n$ such that $v_nw$ is not $G$-invariant.
\end{lem}

\pf Assume  false. If we let $K_n=\ker \mu_n$ then we
get $v\otimes w\in K_n+(V\otimes V)^G$ for each $n,$ so that
$$v\otimes w\in (V\otimes V)^G+\cap_{n\in \Z}K_n.$$
But Lemma 3.1 of [DM2] tells us that $\cap K_n=0,$ whence in fact $v\otimes w
\in (V\otimes V)^G.$ This contradiction proves the lemma. \qed

We turn to the proof of Theorem \ref{t2}, which we prove by induction on
the order of $G.$ Let $W$ be a sub VOA satisfying $V^G\subset W\subset V.$
By Proposition 3.3 of [DM2] it suffices to show that $W=V^H$ for
some subgroup $H$ of $G.$ As in [DM2], $V$ decomposes according to the
simple $G$-modules,
$V=\oplus_{\c\in \Irr(G)}V^{\c}$ and
$W=\oplus_{\c\in \Irr(G)}(W\cap V^{\c}).$

Now if $W=V^G$ there is nothing to prove, so we may assume that $W\cap
V^{\chi}\ne 0$ for some $\chi\in \Irr(G)$ with
$\c\ne 1_G.$ Pick some $0\ne w \in W\cap V^{\c}.$ As $G$ is nilpotent we may
choose a non-identity $g$ in the center of $G.$ Then $g$ acts on $W\cap V^{\c}$
as multiplication by a scalar, say $\l.$ By consideration of $Y(w,z)w$ we see
via Lemma 3.1 of [DM2] that also $W$ contains  nonzero vectors on which $g$
acts as multiplication by $\l^2,$ and similarly $W$ contains nonzero vectors on
which $g$ acts as multiplication by any power of $\l.$

We are going to show that we may assume the existence of a non-identity
element $h$ in the center of $G$ such that if $Z=\<h\>$ then $W^Z\ne V^G.$ If
the scalar $\l$ in the preceding paragraph is equal to 1 we may take
$g=h,$ since in this case $W^Z\supset W\cap V^{\c}\ne 0.$ If $\l\ne 1$ choose,
as we may, some $0 \ne x\in W$ such that $gx=\l^{-1}x.$ Consideration of
$Y(w,z)x$ shows that $g$ fixes each $w_nx,$ so we may as well assume that
each $w_nx$ lies in $V^G.$ Now Lemma \ref{l7.1} shows that
$w\otimes x\in (V\otimes V)^G.$ This can only happen when $\chi$ is a
linear character (i.e., of degree 1), in which case we may chose $h$ to lie in
the kernel of $\c$ if this latter group has order bigger than 1. If it
does not then $G$ is a cyclic group, in which case the theorem has already
been established in [DM2].

So indeed we may assume that $W^Z\ne V^G$ for some $1\ne Z=\<h\><Z(G).$

Now consider the VOA $V^Z.$ It is simple by Theorem 4.4 of [DM2] and
admits $G/Z$ as a faithful automorphism group by Proposition 3.3
of [DM2]. Since $W^Z$ is a sub VOA of $V^Z$ which contains $V^G=(V^Z)^{G/Z}$
then by induction $W^Z=(V^Z)^{H/Z}$ for some subgroup $H/Z<G/Z.$ That
is, $W^Z=V^H.$ Note that since $W^Z\ne V^G$ then $H\ne G$ by Proposition
3.3 (loc. cit.).

So now we are in the situation that $W$ is a sub VOA of $V$ which contains
$V^H$ for some proper subgroup $H<G.$ By induction, $W=V^K$ for some subgroup
$K<H,$ and the theorem is proved. \ \qed

\end{document}